# Detection of domain wall eigenfrequency in infinity-shaped magnetic nanostructures

Mahdi Jamali, Kulothungasagaran Narayanapillai, Jae Hyun Kwon, and Hyunsoo Yang[a]

*Department of Electrical and Computer Engineering, National University of Singapore, 4 Engineering Drive 3, Singapore 117576, Singapore*

The dynamics of a magnetic infinity-shaped nanostructure has been experimentally studied by two different techniques such as the sinusoidal resonance excitation and the damped short pulse excitation to measure the eigenfrequency of domain walls. Direct observation of the magnetic domain wall nucleation has been measured in the frequency domain. Electrical measurements of the domain wall dynamics in the frequency domain reveal the existence of multi-eigenmodes for large excitation amplitudes. The time-resolved measurements show that the frequency of the damped gyration is similar to that of the frequency domain and coexistence of spin wave excitations.

[a] Electronic address: eleyang@nus.edu.sg

Study of the magnetization dynamics in ferromagnetic nanostructures are of great interest among the researchers due to their potential applications for memories[1, 2] and logic devices[3]. Two general techniques have been developed to characterize domain wall dynamics such as time-resolved X-ray microscopy[4-6] and electrical methods.[7-10] Recent investigations of the domain wall dynamics in confined magnetic structures such as a disk and an infinity-shaped (∞) nanostructure have shown a fascinating behavior including ultrafast dynamics of the magnetization[11-14]. While there have been some experimental reports on the measurement of the domain wall dynamics in an infinity-shaped nanostructure using the X-ray microscopy method[15, 16] in the past few years, there has been little study on the electrical measurement of domain wall eigenfrequency in this structure.

In this letter, we present an all-electrical measurement of magnetic dynamics in an infinity-shaped nanostructure using two different techniques, such as the rectification method in the frequency domain and the pulse excitation method in the time domain in order to measure the eigenmode frequencies and damping factor. It is found that by using the domain wall rectification effect,[8, 17, 18] it is possible to detect the nucleation of the domain wall in the presence of a perpendicular magnetic field. Multiple eigenmodes have been observed in the frequency spectrum of the magnetic domain structure. Furthermore, it is found that the resonance frequency of the domain wall depends weakly on the in-plane external magnetic field at low bias fields, and we observe a sizable shift in the resonance frequency at high in-plane magnetic fields. By increasing the excitation amplitude, higher frequency modes are detected. The time-resolved response to a narrow voltage pulse (< 5 ns) has been performed. The eigenfrequency of the damped gyration is found to be similar to the resonance frequency measured by the rectification method in the frequency domain. Furthermore, the dynamics of domain walls coexists with spin wave excitations inside the ferromagnetic structure.

The infinity-shaped ferromagnetic nanostructure is shown in Fig. 1, which is a top-view scanning electron microscope (SEM) image. Devices are fabricated by dc sputter deposition of Ta (2 nm)/Ni$_{81}$Fe$_{19}$ (20 nm)/Ta (2 nm) on a Si/SiO$_2$ (300 nm) substrate, followed by the patterning of the infinity-shaped nanostructure using e-beam lithography and Ar ion milling. After ion milling, 2 nm SiO$_2$ is rf sputter deposited without breaking the vacuum to avoid the oxidation of the ferromagnetic nanostructure. A second e-beam lithography step is used to pattern Ta (5 nm)/Cu (100 nm) contacts. Before contact deposition, the top surface of the ferromagnetic structure is etched through 2 nm to clean the interface.

In order to measure magnetization dynamics, we have used the domain wall rectification effect, and the resonance frequency of the magnetic structure has been measured using a homodyne circuit as shown in Fig. 1. A low frequency voltage from a lock-in amplifier is mixed with a high frequency sinusoidal signal from a microwave signal generator using an amplitude modulator (AM) in order to excite the magnetic domain wall dynamics. The resultant signal is fed to the device through A$_1$A$_2$ contacts. The voltage between B$_1$B$_2$ contacts is measured by the lock-in amplifier which locks on the low frequency component. Assuming that the lock-in frequency ($\omega_b$) is much smaller than the signal generator frequency ($\omega_c$), (i.e. $\omega_c \gg \omega_b$), the output signal of the amplitude modulator can be written as $J_{c0}\sin(\omega_c t + \phi_1)[1 + m\sin(\omega_b t)]$, where $m$ is the amplitude modulation constant equals to the ratio of the lock-in to signal generator amplitude, and $J_{c0}$ is the current density. When the frequency of the signal generator is close to the resonance frequency of the magnetic structure, the magnetic structure exhibits a resonant gyrotropic motion, and the resistance across the magnetic domain wall in the nanostructure would contain an oscillatory component due to the anisotropic magnetoresistance (AMR) effect.[7, 9] The resultant changes in the resistance could be written as $R_0 \sin(\omega_c t + \phi_2)[J_{c0}\sin(\omega_c t + \phi_1)\{1 + m\sin(\omega_b t)\}]$, where $R_0$ is the change in the resistance due to the gyrotropic motion of the magnetic structure that could be only excited by the high frequency component of the input signal. The amplitude of the resultant sinusoidal voltage having a frequency $\omega_b$, which is measured by the lock-in amplifier, is proportional to $0.5mR_0 J_{c0} \cos(\phi_2 - \phi_1)$, which is the phase difference between the input signal and the change in the device resistance.[9] The frequency of the signal generated by the lock-in amplifier is set to 931.7 Hz and the signal generator frequency is swept from 10 MHz to 1 GHz in 0.5 MHz increments.

In order to nucleate magnetic domain walls in the structure, a perpendicular magnetic field (z-direction) is applied to the nanostructure and the lock-in voltage is monitored at different frequencies. As can be seen in Fig. 2(a), above certain values of the magnetic field ($H_c$ = 1275 Oe), a distinct peak around 73.5 MHz appears at the output signal which was not present at lower fields. Furthermore, by increasing the magnetic field above $H_c$, the peak position remains invariant. After the external field has been turned off, we performed the measurements again. It is found that the position of the peaks remain the same in the frequency domain. After nucleation of domain walls, the magnetic force microscopy (MFM) data confirm the presence of magnetic domain walls as shown in Fig. 2(b).

In order to see the effect of the amplitude of lock-in voltage on the output spectrum of the magnetic structure, the lock-in voltage has been changed from 100 to 500 mV for a fixed signal generator voltage of 500 mV corresponding to $m$ = 0.2 up to $m$ = 1. As can be seen in Fig. 2(c), the output spectrum is almost independent of lock-in voltages, which corresponds to our expectation that only the high frequency component of the input signal should excite the dynamics of inside the magnetic structure. In addition, two more peaks are observed at 117.5 and 166.5 MHz, which could be either due to the different magnetic structures or due to the nonlinear eigenmodes of the magnetic structure that have been previously reported[19, 20].

Increasing the amplitude of the signal generator increases the input current density and it could ideally excite higher eigenfrequencies of the magnetic structure. The responses of the magnetic structure for different signal generator amplitudes with a fixed *m* of 0.95 are shown in Fig. 2(d). For small signal generator amplitudes, the current density is too low to excite dynamics inside the magnetic structure. For a signal generator voltage of 500 mV, the current density in the branches of the nanostructure is approximately $3\times10^7$ A/cm$^2$. The second and third peaks above the main peak at 73.5 MHz are observable for signal generator voltages greater than 300 mV. Similar nonlinear dynamics for the vortex structure have been reported by the injection of a high current density.[21]

Measurements of the resonance frequency of the magnetic structure have been performed on nanostructures with three different sizes of *D*, specifically 410, 480, and 720 nm. The frequency spectrum of each device at zero magnetic field is shown in Fig. 3(a). The main resonance frequencies are found to be 176.4, 112.3, and 73.5 MHz in the devices whose lengths of *D* are 410, 480, and 720 nm, respectively [Fig. 3(b)]. A decrease in the device size results in an increase in the resonance frequency, which is similar to results of the magnetic vortices in a circular nanodot that was previously reported.[7]

The effect of the magnetic field on the dynamics of the magnetic structure has been also studied. The magnetic field is applied in the *x*-direction, and the gyration dynamics of the domain wall is measured by sweeping the frequency of the signal generator, delivering power at an amplitude of 500 mV as can be seen in Fig. 3(c). It is found that the output spectrum of the magnetic structure do not change with the magnetic field up to 570 Oe. These results are in line with those of previous magnetic vortex studies in a ferromagnetic dot.[22, 23] Upon application of a magnetic field above 570 Oe, the frequency spectrum of the magnetic structure changes and the position of the main peak shifts from 73.5 to 247.5 MHz. Furthermore, the amount of changes ($\Delta V$) in the amplitude of the output signal at the resonance frequency varies from around 11 to 13.6 µV. The $\Delta V$ is related to the gyrotropic amplitude of the antivortex/vortex structure and the AMR gradient across the magnetic structure,[9] and thus a 23% variation in $\Delta V$ from 11 µV to 13.6 µV can be explained by the transformation of the magnetic structure configuration to another magnetic structure.[24]

In order to better understand this phenomenon, the magnetic field is removed, and the eigenfrequency of the magnetic structure is measured again as shown in Fig. 3(d). It is found that the new magnetic structure preserves its structure and the main mode at 247.5 MHz is still present. Moreover, any significant change in the resonance frequency of the magnetic structure has not been observed even with the application of a magnetic field up to 2 kOe in the $x$-direction. Therefore, we conclude that the new configuration of the magnetic structure is in a stable state. Micromagnetic simulations have been performed to find the new magnetic structure as shown in the inset of Fig. 3(d). Furthermore, MFM imaging of the magnetic domain structure also reveals the change in the magnetic domain structures as can be seen in the inset of Fig. 3(b) compared to the original magnetic domain structure in Fig. 2(b).

The vortex or antivortex resonance frequency may be written as $f = k_M/2\pi G_0$,[25] where $k_M$ is the stiffness of magnetic structures due to the displacement of the core from its equilibrium position and $G_0$ is the gyration vector amplitude. The stiffness is related to the magnetic susceptibility $\chi_M = dM_x/dH_x$ by $k_M = \pi L M_s^2 \xi^2 / \chi_M$,[23, 25] where $\xi$ is a parameter which describes the type of the boundary condition ($\xi \approx 1$).[23] To calculate $\chi_M$, $1/\chi_M = 2\beta[\ln(8/\beta)-0.5]$ has been used,[23] where $\beta (= L/R)$ is the aspect ratio of the nanostructure thickness ($L$) over radius ($R$). By using a $D$ parameter of 720 nm as defined in Fig. 3 (b) and the properties of Permalloy ($M_s = 8\times10^5$ A/m and $\gamma = 1.76\times10^2$ GHz/T), a resonance frequency of 83 MHz is calculated, which is in good agreement with the main peak of 73.5 MHz in the present experimental result.

In addition to the frequency domain studies, we have investigated the transient response of domain walls in the same nanostructure, by utilizing the circuit configuration as shown in Fig. 4(a). A short voltage pulse is applied between $A_1A_2$ ports to displace the domain wall from its equilibrium position and excite dynamics inside the magnetic domain walls. The voltage pulse has an 80 ps rise and fall time, and the pulse width is 5 ns. After the excitation has removed, the gyrotropic motion of domain walls remains and can be detected by applying a dc current through a bias tee in the excitation port, which has been also proposed previously[10]. The dc current (50 μA) with a current density $\sim 6\times10^6$ A/cm$^2$ is sufficiently small in comparison to the excitation current density and does not affect the domain wall dynamics. The output signal is measured by a Tektronix 6 GHz real time oscilloscope between $B_1B_2$ ports. The voltage pulses have a repetition of 1 kHz and the output signal is averaged 10000 times in the oscilloscope to improve the signal-to-noise ratio. Figure 4(b) shows the output signal for an excitation pulse amplitude of 4 V ($J=1.2\times10^8$ A/cm$^2$). Close to the excitation pulse, the output signal is complex and contains high frequency components, while the behavior of the output signal is similar to a damped sinusoidal away from the excitation. The input excitation has been overlaid in Fig. 4(b) to show that the high-frequency components exist mostly in the first 30 ns of the transient response. At the rise and fall time of the input pulse, there are sharp changes at the output signal. Right after the rising and falling edges, there are some high frequency components that decay very quickly, and only low frequency components remain after 30 ns which have originated from the domain wall damped motion. The high frequency dynamics less than 30 ns could be associated to the spin wave excitation often observed in the time-resolved spin wave measurement by a voltage pulse.[26,27]

A curve fitting has been performed on the output signal to calculate the damping ratio and damped resonance frequency of the motion. The formula used in the curve fitting is $V = Ae^{-\Gamma t}\sin(2\pi ft + \Phi) + V_0$, with fitting parameters $A$ = 0.0158 V, $\Gamma$ = 3.02×10$^7$ Hz, $f$ = 70.36 MHz, $\Phi$ = 0.539 radian, and $V_0 \sim$ 0 V. The damped resonance frequency is found to be 70.36 MHz, which is very close to the resonance frequency measured by the homodyne technique (73.5 MHz) in Fig. 2. The induced voltage due to the domain wall damped motion has a decay time (1/$\Gamma$) of about 33 ns, which is comparable to the theoretical calculation for an abrupt current pulse excitation[10].

In conclusion, the dynamics of an infinity-shaped magnetic nanostructure has been studied experimentally using two different electrical methods, such as resonance sinusoidal and transient pulse excitations. We demonstrate that the domain nucleation field could be measured which was found to be about 1275 Oe in our structure. Furthermore, multiple eigenfrequencies have been observed in the frequency response of the magnetic structure for large excitation amplitudes. For the pulse excitation of the magnetic domain wall, the damped frequency of transient response is very close to the main eigenfrequency of the domain wall measured by resonance excitation. Furthermore, the transient response of the magnetic domain wall accompanies spin wave excitation in the ferromagnetic nanostructure. Our demonstration of all-electrical measurements of magnetic domain wall dynamics paves the way for a better understanding of magnetization dynamics in various magnetic nanostructures.

This work is supported by the Singapore National Research Foundation under CRP Award No. NRF-CRP 4-2008-06.

Figure Captions

FIG. 1. A SEM image of the ferromagnetic structure with a schematic representation of the electric circuit used for the measurement of the domain wall resonance frequency.

FIG. 2. (a) The frequency spectra of the magnetic structure for different values of the perpendicular magnetic field with a 3 µV voltage offset for each data set. (b) A MFM image of the magnetic domain walls after nucleation. (c) Frequency spectra of the domain walls for different lock-in amplifier voltages normalized by $m$ with a 6 µV voltage offset for each data set. (d) The frequency spectrum of the domain walls for different values of the signal generator amplitudes on a logarithmic scale. All the data are from a device of $D = 720$ nm.

FIG. 3. (a) The frequency spectrum of the magnetic domain structures for different device sizes. (b) The resonance frequency versus the device size, $D$, which is defined in the inset of (b). (c) The effect of the in-plane magnetic field in the $x$-direction on the frequency spectra with a 12 µV voltage offset for each data set. (d) The frequency response of the new magnetic structure at different magnetic fields in the $x$-direction. A MFM image is shown in the inset of (b) after the new magnetic structure is formed. The micromagnetic simulation of the new magnetic structure is shown in the inset of (d).

FIG. 4. (a) The electric circuit configuration for the measurement of transient response. (b) The measured output signal with the corresponding excitation pulse and the curve fitting data. The data are from a device of $D = 720$ nm.

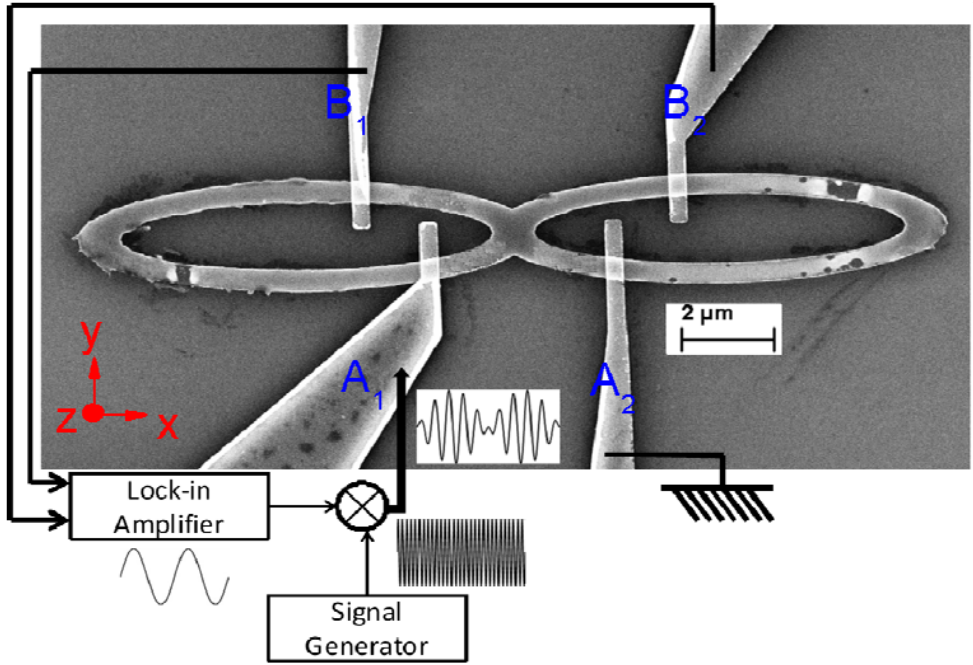

Figure 1.

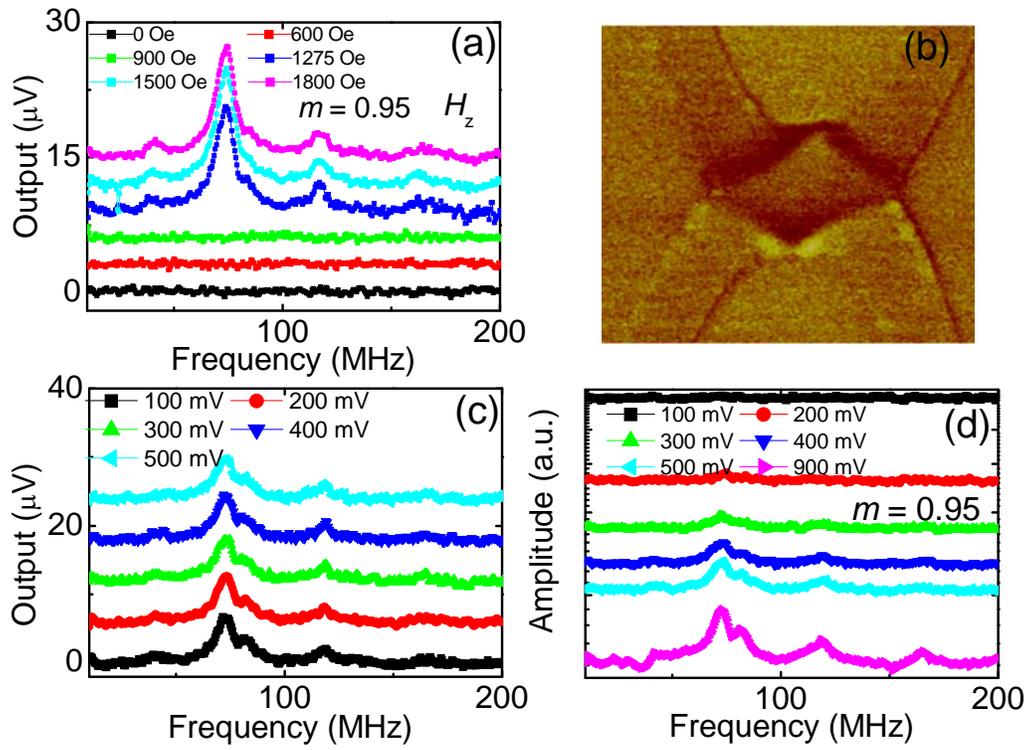

Figure 2.

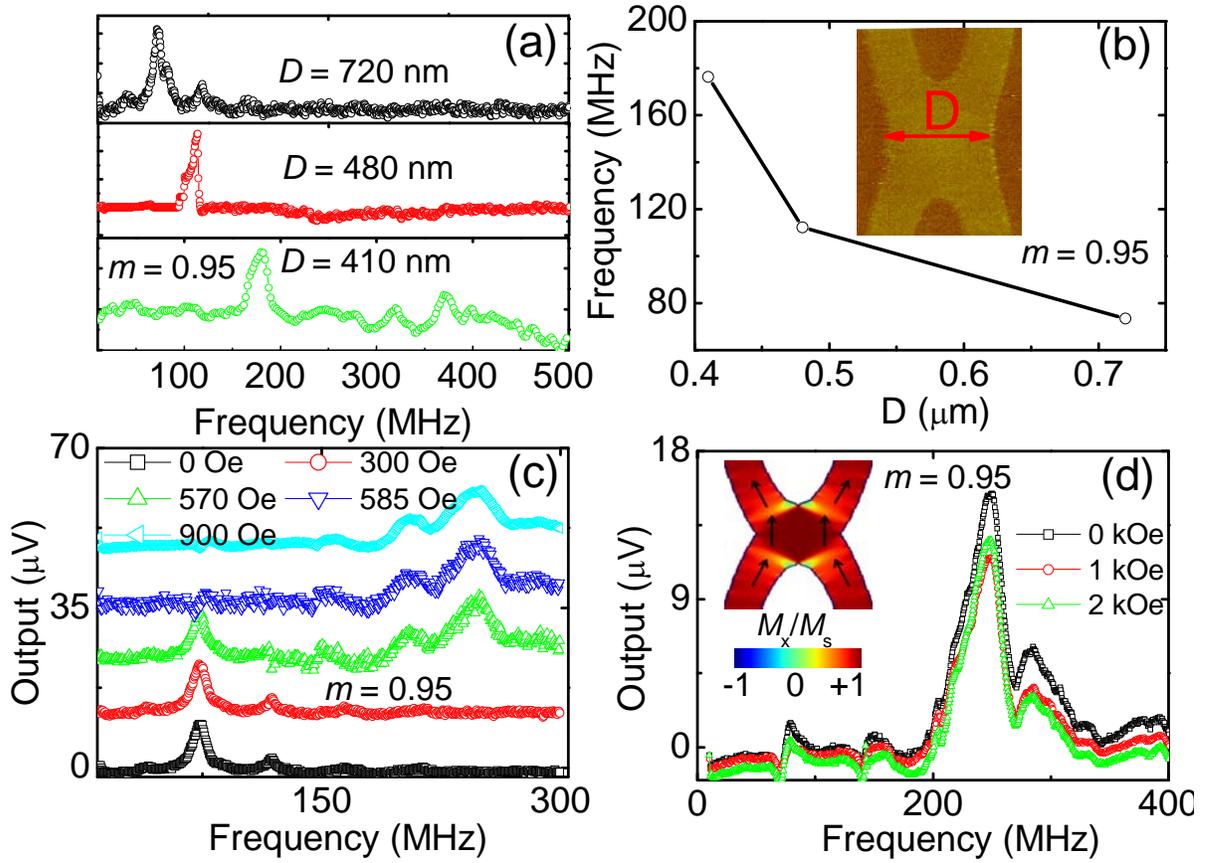

Figure 3.

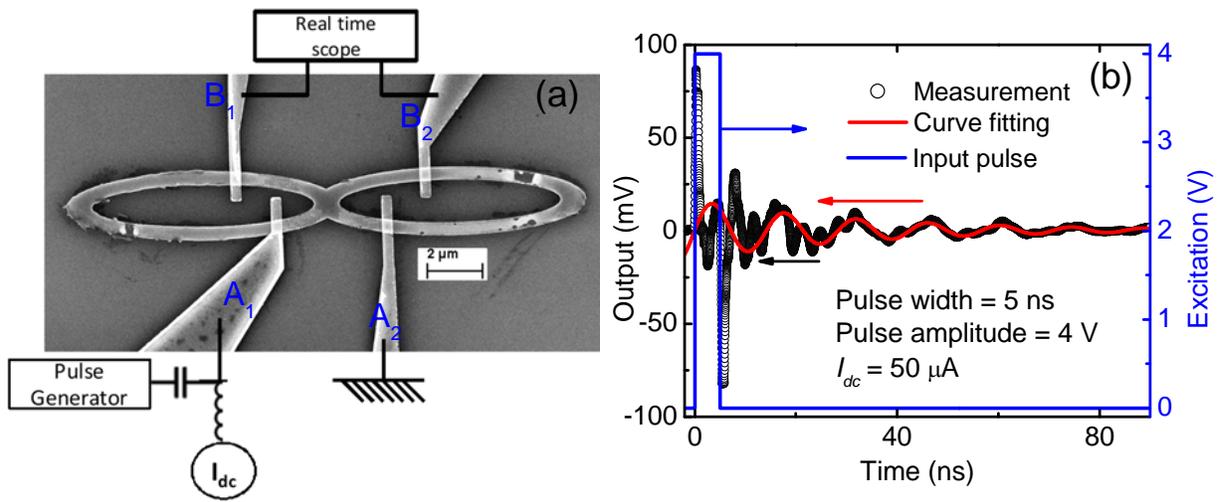

Figure 4